\documentclass{PoS}

\newcommand{\rT}{{\mathrm{T}}}

\newcommand{\LO}{{\mathrm{LO}}}
\newcommand{\NLO}{{\mathrm{NLO}}}
\newcommand{\EW}{{\mathrm{EW}}}

\def\mathswitchr#1{\relax\ifmmode{\mathrm{#1}}\else$\mathrm{#1}$\fi}

\newcommand{\Pe}{\mathswitchr e}

\newcommand{\Pp}{\mathswitchr p}
\newcommand{\Pj}{\mathswitchr j}

\newcommand{\PW}{\mathswitchr W}

\newcommand{\Pl}{\ell}

\def\mathswitch#1{\relax\ifmmode#1\else$#1$\fi}

\newcommand{\MW}{\mathswitch {M_\PW}}

\newcommand{\TeV}{\unskip\,\mathrm{TeV}}
\newcommand{\GeV}{\unskip\,\mathrm{GeV}}

\title{NLO electroweak corrections to vector-boson scattering at the LHC}

\ShortTitle{NLO electroweak corrections to vector-boson scattering at the LHC}

\author{Benedikt Biedermann, Ansgar Denner, \speaker{Mathieu Pellen} \\
        Universit\"at W\"urzburg, %
        Institut f\"ur Theoretische Physik und Astrophysik, %
        Emil-Hilb-Weg 22,
        97074 W\"urzburg, %
        Germany \\
        E-mail: \email{mpellen@physik.uni-wuerzburg.de}}

\abstract{
Recently, a measurement of the vector-boson scattering process with same-sign W bosons has been reported by the CMS collaboration.
Hence it is of prime importance to have precise predictions with next-to-leading order (NLO) accuracy.
In these proceedings, we report on a recent NLO electroweak computation to the full process $\Pp\Pp\to\mu^+\nu_\mu\Pe^+\nu_{\Pe}\Pj\Pj$.
As realistic experimental event selections are applied to the final state, it can directly be compared with experimental measurements.
This is particularly important as the corrections turn out to be surprisingly large and even exceed the NLO QCD corrections.
The NLO electroweak predictions are presented at the cross-section and differential distribution level.
}

\FullConference{XXV International Workshop on Deep-Inelastic Scattering and Related Subjects\\
		3-7 April 2017\\
		University of Birmingham, UK}

\begin{document}

\section{Introduction}

During run~I of the Large Hadron Collider (LHC), evidence for the existence of the vector-boson scattering (VBS) process with two positively charged W bosons has been found \cite{Aad:2014zda,Aaboud:2016ffv,Khachatryan:2014sta}.
For run~II, the CMS collaboration has already reported a measurement \cite{CMS:2017adb}.
Precise and appropriate predictions for this process are thus very relevant.
In this context \emph{precise} means next-to-leading order (NLO) accurate and \emph{appropriate} means that the simulations are done with realistic experimental cuts.
This implies that the NLO QCD corrections but also the electroweak (EW) ones should be considered.
The former are already available \cite{Jager:2009xx,Jager:2011ms,Denner:2012dz,Baglio:2014uba,Rauch:2016pai} while the latter have been computed only recently \cite{Biedermann:2016yds} and are reported in these proceedings.
The NLO EW corrections are unusually large for EW corrections at the LHC, and even exceed the NLO QCD corrections \cite{Biedermann:2017bss}.
The calculation presented here contains all possible off-shell and non-resonant effects as it describes the full $\Pp\Pp\to\mu^+\nu_\mu\Pe^+\nu_{\Pe}\Pj\Pj$ process.
We consider the leading order (LO) at order $\mathcal{O}{\left(\alpha^{6}\right)}$, while the NLO EW corrections are defined at order $\mathcal{O}{\left(\alpha^{7}\right)}$.
Finally, realistic experimental cuts are applied, allowing for a direct comparison with the experimental measurements.

\section{Event selection}

The predictions presented here are for the LHC running at the center-of-mass energy of $13\TeV$.
The input parameters can be found in Ref.~\cite{Biedermann:2016yds}.
The event selection is adopted to the one used in experiments to single out the VBS process over its irreducible background \cite{Aad:2014zda,Aaboud:2016ffv,Khachatryan:2014sta,CMS:2017adb}.
To recombine the photons with charged particles, the anti-$k_\rT$ algorithm \cite{Cacciari:2008gp} with $R=0.1$ is used.
The experimental signature of the process is characterised by two leptons of positive charge, two QCD jets, and missing transverse energy.
Each charged lepton and jet has to fulfil the following requirements for their transverse momentum and rapidity:
\begin{eqnarray}
 p_{\rT,\ell} &{}>& 20\GeV, \quad |y_{\ell}| < 2.5, \quad  \Delta R_{\Pl\Pl} > 0.3,\\
 p_{\rT,\rm j} &{}>& 30\GeV, \quad |y_{\rm j}| < 4.5, \quad \Delta R_{\Pj\Pl} > 0.3.
\end{eqnarray}
In addition, the missing transverse energy has to be larger than $40\GeV$.
For the pair of jets, the typical VBS event selections are applied.
These comprise an invariant-mass cut and a cut on the difference of the rapidities,
\begin{equation}
 M_{\Pj \Pj} > 500\GeV, \quad |\Delta y_{\Pj \Pj}| > 2.5.
\end{equation}

\section{Numerical results}

The numerical results have been obtained from two different private Monte Carlo
programs which have been used to compute NLO QCD and EW corrections to high-multiplicity processes.
To evaluate all tree and one-loop amplitudes, the computer code {\sc Recola}~\cite{Actis:2012qn,Actis:2016mpe} and the {\sc Collier}
library \cite{Denner:2014gla,Denner:2016kdg} have been used. 
The infrared singularities are handled via the Catani--Seymour dipole subtraction
formalism \cite{Catani:1996vz,Dittmaier:1999mb}.
To treat unstable particles, we rely on the the complex-mass scheme \cite{Denner:1999gp,Denner:2005fg}.
Finally, regarding the electromagnetic coupling, the $G_\mu$ scheme \cite{Denner:2000bj} is utilised.

\begin{table}
\begin{center}
\begin{tabular}
{|ccc|}
\hline
$\sigma^{\LO}$~[fb] &  $\sigma^{\NLO}_{\EW}$~[fb] & $\delta_{\EW}~[\%]$ 
\\
\hline
$\phantom{1}1.5348(2)$ & $\phantom{1}1.2895(6)$& $-16.0$ 
\\
\hline
\end{tabular}
\end{center}
\vspace{-0.2cm}
\caption{Fiducial cross sections at LO and NLO EW accuracy for $\Pp \Pp \to \mu^+ \nu_\mu \Pe^+ \nu_{\Pe} \Pj\Pj$ expressed in femtobarn.
The digit in parenthesis represents the integration error.
The relative EW corrections $\delta_{\EW}$ are given in per cent.}
\label{table}
\end{table}

The numerical results obtained are particularly interesting.
As it can be seen from Table~\ref{table}, the EW corrections are already large for the fiducial cross section.
They reach $-16\%$ and thus are even larger than the QCD ones \cite{Biedermann:2017bss}.
This is surprising as usually the EW corrections are mainly driven by EW Sudakov logarithms that grow negatively large only in the tail of the distributions (which are themselves suppressed).

\begin{figure}[h!]
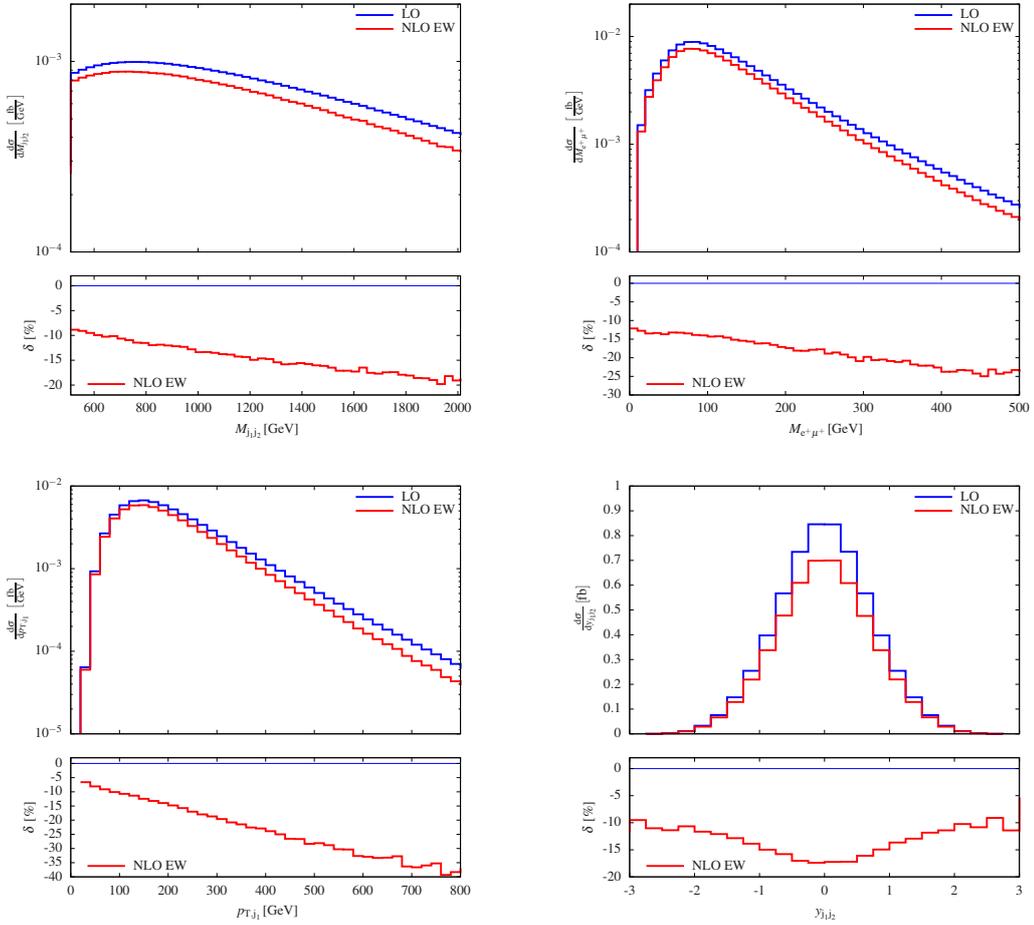

\begin{center}
\includegraphics[width=.48\textwidth]{{{histogram_invariant_mass_mjj12}}}
\includegraphics[width=.48\textwidth]{{{histogram_invariant_mass_truth_epmu}}}
\includegraphics[width=.48\textwidth]{{{histogram_transverse_momentum_j1}}}
\includegraphics[width=.48\textwidth]{{{histogram_rapidity_j1j2}}}
\end{center}
\vspace{-0.7cm}
\caption{Differential distributions for $\Pp\Pp\to\mu^+\nu_\mu\Pe^+\nu_{\Pe}\Pj\Pj$
including NLO EW corrections (upper panels) and relative NLO EW corrections (lower panels):
dijet invariant-mass distribution (top left),
invariant-mass distribution of the positron--anti-muon system (top right),
transverse-momentum distribution of the hardest jet (bottom left), and
rapidity distribution of the leading jet pair (bottom right).}
\label{figures}
\end{figure}

The effects of Sudakov logarithms are clearly visible in the distributions in the invariant mass (both for the two charged leptons and the two jets) and in the distribution in the transverse-momentum of the hardest jet.\footnote{The hardest jet is determined according to transverse-momenta ordering.}
These are shown in Fig.~\ref{figures}, where the LO and NLO EW predictions are displayed (upper panels) as well as the relative EW corrections (lower panels) in per cent.
The corrections reach $-40\%$ for the transverse momentum of the leading jet at $800\GeV$.
For the distribution in the rapidity of the two jets, the EW corrections are large for the typical VBS configuration where the two jets are back to back (corresponding to small $y_{\Pj_1 \Pj_2}$).
When going away from this configuration (larger $y_{\Pj_1 \Pj_2}$) the magnitude of the corrections decreases.

The origin of the large EW corrections has been identified in Ref.~\cite{Biedermann:2016yds}.
These are driven by the bosonic part of the virtual contributions.
Studying $\PW^+ \PW^+ \to \PW^+ \PW^+$ scattering in a leading-logarithmic approximation \cite{Denner:2000jv} gives a very good estimate of the corrections.
The obtained formula depends only on the EW Casimir operator and the scale of the process.
It explains why the corrections are 3-4 times bigger than for $q\bar{q} \to \PW^+ \PW^+$.
Indeed, the Casimir operators are larger for vector bosons than for fermions and the scale $Q = \langle m_{\rm 4\Pl}\rangle \sim 390\GeV$ entering the Sudakov logarithm $\log \left(Q^2 / \MW^2 \right)$ is also larger as the scattering features a massive $t$-channel exchange.
This reveals that the large EW corrections observed are an intrinsic feature of the VBS process at the LHC.

\section{Conclusion}

In these proceedings, the off-shell computation at NLO EW accuracy of the VBS for two same-sign W bosons has been presented \cite{Biedermann:2016yds}.
Besides being a theoretical and numerical challenge, it also constitutes a relevant piece of information for the experimental collaborations in their quest to measure VBS processes precisely.
Indeed, the calculation features all non-resonant and off-shell effects with a realistic final state.
In addition, typical experimental event selections are applied which allows one to compare these predictions directly to the experimental measurements.
The fact that these corrections are large (even larger than the QCD ones \cite{Biedermann:2017bss}) renders them even more relevant for the experimental analysis.

\acknowledgments
We thank Jean-Nicolas Lang and Sandro Uccirati for
supporting {\sc Recola} and Robert Feger
for assistance with {\sc MoCaNLO}.
We acknowledge financial support by the German Federal Ministry for Education and
Research (BMBF) under contract no.~05H15WWCA1 and the German Science
Foundation (DFG) under reference number DE 623/6-1.

\end{document}